\documentclass[twocolumn,showpacs,preprintnumbers,amsmath,amssymb]{revtex4}
\usepackage{graphicx}% Include figure files
\usepackage{dcolumn}% Align table columns on decimal point
\usepackage{bm}% bold math

\newcommand{\be}{\begin{equation}} 
\newcommand{\en}{\end{equation}}
\newcommand{\bea}{\begin{eqnarray}}
\newcommand{\ena}{\end{eqnarray}}

\newcommand{\hbo}{\hbox to 1 true cm {\hfill } } 
\newcommand{\tr}{\hbox{tr}}

\begin{document}

\preprint{ UNITUE-THEP/04-2004}

\title{ A novel improved action for SU(3) lattice gauge theory } 
% Force line breaks with \\

\author{Kurt Langfeld}
 \email{kurt.langfeld@uni-tuebingen.de} 

\affiliation{%
Insitut f\"ur Theoretische Physik, Universit\"at T\"ubingen \\ 
Auf der Morgenstelle 14, D-72076 T\"ubingen, Germany. \\ 
}%

\date{(revised) Juli 22, 2004 }% It is always \today, today,

\begin{abstract}
SU(3) lattice gauge theory is studied by means of an improved 
action where a $2 \times 2$ Wilson loop is supplemented to the 
standard plaquette term. By contrast to earlier studies using 
a tree level improvement, the prefactor of the $2 \times 2$ Wilson term 
is determined by minimizing the breaking of rotational symmetry 
detected from the static quark-antiquark potential. On coarse lattices, 
the novel action is superior to the Iwasaki action  and comparable with 
DBW2 action. The scaling behavior of the novel action is studied 
by using the static quark potential and the ratio of the 
deconfinement temperature and the string tension. 
\end{abstract}

%Use showkeys class option if keyword
\maketitle

A systematic approach to the non-perturbative regime of Yang-Mills 
theory is provided by computer simulations of lattice gauge theory. 
Regularization is introduced by replacing continuous spacetime 
by a hypercubic lattice with lattice spacing $a$. Improved actions for 
those lattice gauge simulations have attracted much 
interest in the recent past, since they allow for simulations on coarse 
lattices without an overwhelming impact of discretization 
errors. For example if the infra red regime of QCD Green functions 
is addressed (see e.g.~\cite{Bonnet:2000kw,Bloch:2003sk}), the low lying 
momenta are of order $1/Na$, where $N$ the 
number of lattice points in one direction. For a reasonable amount of 
lattice points, large lattice spacings are highly desirable for 
these purposes. Using the standard Wilson action, large lattice spacings 
give rise to sizable violations of rotational symmetry. These 
induce a systematic error to the lattice data and severely limit their 
significance. The benefit of improved actions is that they make use 
of large lattice spacings without sacrificing the continuum limit. 

\vskip 0.2cm
In lattice gauge theory, 
dynamical degrees of freedom are unitary matrices $U_\mu (x) \in SU(3)$
associated with the links of the lattice. 
The partition function is defined by 
$$ 
Z \; = \; \int {\cal D}U_\mu \; \exp \bigl\{ -  S_{\mathrm{latt}} [U] 
\bigr\} \; . 
$$ 
In order to recover continuum Yang-Mills theory, one 
demands in the limit $a\rightarrow 0$ (adopting minimal Landau gauge) 
\bea 
U_\mu (x) &=& \exp \bigl\{ i t^b \, A^b_\mu (x) a \bigr\} \; , 
\\ 
S_{\mathrm{latt}} &=&  \sum _{x,\mu \nu } \bigl[ 
\frac{1}{4} F_{\mu \nu }^b F_{\mu \nu }^b \, a^4 \; + \; O_1 \, a^6 
+ \ldots \bigr] \; , 
\label{eq:o1} 
\ena 
where $t^a$ are the generators of the SU(3) group, $A^b_\mu(x)$ are 
the gauge fields and 
\begin{table}[thb]
\begin{ruledtabular}
\caption{ \label{tab:1}  Coupling constants of the ``$1\times2$'' term 
   contributing to the improved action. } 
\begin{tabular}{ccccc}
  & Wilson & Tadpole improved  & Iwasaki  & DBW2 \\ \hline 
 $f$ & $0$ & $1/20 u_0^2$ & $0.0907$ & $0.1148$ \\ 
\end{tabular}
\end{ruledtabular}
\end{table}
$ F_{\mu \nu }^b$ is the usual non-Abelian field strength tensor. 
When one designs a realization of $S_{\mathrm{latt}}$ in terms of the 
matrices $U_\mu (x)$, it is essential that $S_{\mathrm{latt}}$ 
is gauge invariant for any choice of the lattice spacing $a$, 
thereby enforcing gauge invariance in the continuum limit. 
The most simple choice for such an action  is the 
Wilson action:
\be 
S_{\mathrm{latt}}^{wil} \; = \; \beta \; \sum _{x, \mu > \nu } 
\frac{1}{3} \mathrm{Re} 
\, \tr \bigl\{ 1 - W^{1 \times 1 }_{\mu \nu }(x)  \bigr\} \; , 
\label{eq:wil}
\en 
where $W^{n \times m }_{\mu \nu }(x)$ is the ${n \times m }$ rectangular 
Wilson loop, located at $x$, with extension $n$ in $\mu $ and 
extension $m$ in $\nu $ direction. 
$\beta $ is related to the bare gauge coupling $g_0$ by 
$\beta = 6/g_0^2$. 
Universality implies that any other lattice action which 
meets these requirements also serves the purpose in the limit 
$a \rightarrow 0$. The crucial point, however, is that 
computer simulations are carried out on finite lattices consisting of 
$N^4$ lattice points. The corresponding extension of the lattice 
in one particular direction is given by $L=N a$. If we insist on 
$ L > 1.2 \, $fm, the lattice spacing must not be smaller than $a = 
0.075 \,$fm for a $16^4$ grid. The central idea is to construct 
a lattice action which minimizes the contribution of the 
irrelevant terms such as $O_1$ in (\ref{eq:o1}). At the expense 
of a more complicated lattice action, simulations might be 
performed on coarse lattices and might produce results with 
little corrections even for sizable values $a$. 
Several proposals have been made 
for such actions: tadpole improved actions~\cite{Symanzik:1983dc}, 
improved actions obtained by lattice perturbation 
theory~\cite{Luscher:1985zq,Alford:1995hw}, improved actions based 
on renormalization group studies~\cite{Iwasaki:1983ck,deForcrand:1999bi}, 
and ``perfect''actions operating close to the fixed point 
of the theory~\cite{DeGrand:1995jk,Hasenfratz:1997ft}.

\begin{figure}[thb]
\includegraphics[width=7cm]{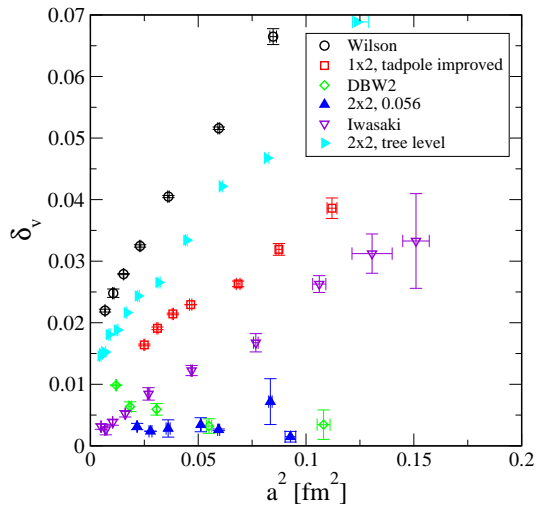} 
\caption{\label{fig:1} The measure $\delta _v$ of the breaking 
of rotational symmetry for several improved actions as 
function of the lattice spacing $a$ in physical units.
}
\end{figure}
\vskip 0.2cm 
Let us focus on improved actions of the type 
\bea 
S_{\mathrm{latt}}^{imp} &=& \beta \; \sum _{x, \mu > \nu } 
\biggl[ \frac{1}{3} \mathrm{Re} 
\, \tr \bigl\{ 1 - W^{1 \times 1 }_{\mu \nu }(x)  \bigr\} 
\nonumber \\ 
&-& f \, 
\frac{1}{3} \mathrm{Re} \, \tr \, \bigl\{ 2 - W^{1 \times 2 }_{\mu \nu }(x) 
- W^{2 \times 1 }_{\mu \nu }(x)  \bigr\} \; \biggr] \; . 
\label{eq:imp}
\ena 
Different ideas on improvement result in different proposals for 
$f$. More than ten years ago, Iwasaki used perturbation theory for 
a renormalization group analysis and suggested $f=0.0907$. 
The so-called DBW2 action,  derived from ``double blocked Wilson'' 
configurations~\cite{deForcrand:1999bi}, chooses $f=0.1148$. 
The ``tadpole improved tree level'' action~\cite{Symanzik:1983dc} 
(see also \cite{Bonnet:2001rc} for details) favors a $\beta $ 
dependent factor $f = 1/20 \, u_0^2 $, where 
$$ 
u_0 (\beta ) \; = \; \left[ \frac{1}{3} \mathrm{Re} \, 
\left\langle \tr W^{1 \times 1 }_{\mu \nu }(x) \right\rangle
\right]^{1/4} 
$$
must be self-consistently determined. See Table \ref{tab:1} for a 
summary. A thorough study of the scaling behavior of several 
``$1 \times 2$'' actions was recently performed by 
Necco~\cite{Necco:2003vh}. It was found at least for small $a$ that 
the DBW2 action shows larger lattice artefacts than the 
Iwasaki action. 

\begin{table}[thb]
\begin{ruledtabular}
\caption{ \label{tab:4} String tension in units of the lattice spacing, 
$\sigma  a^2$, using the novel action with $f=0.056$. 
} 
\begin{tabular}{cccc}
  $\beta $ & 6.0 & 6.2 & 6.4  \\ \hline 
  $\sigma a^2 $ & 0.343(3) & 0.247(1) & 0.178(10) 
\end{tabular}
\end{ruledtabular}
\end{table}
\begin{figure}[bht]
\includegraphics[width=7cm]{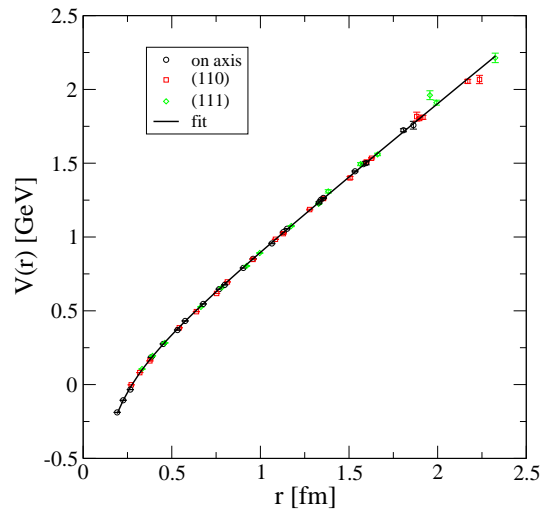} 
\caption{\label{fig:2} The static quark potential for the novel 
action ($f=0.056$) for a $16^4$ lattice. 
}
\end{figure}
\vskip 0.2cm 
Here, we propose a new type of improvement which is based upon the 
lattice action of type 
\bea 
S_{\mathrm{latt}}^{new} &=& \beta \; \sum _{x, \mu > \nu } 
\biggl[ \frac{1}{3} \mathrm{Re} 
\, \tr \bigl\{ 1 - W^{1 \times 1 }_{\mu \nu }(x)  \bigr\} 
\nonumber \\
&-& f \, \frac{1}{3} \mathrm{Re} \, \tr \, 
\bigl\{ 1 - W^{2 \times 2 }_{\mu \nu }(x) 
\; \bigr\}\ \; \biggr] \; . 
\label{eq:new}
\ena 
Since $W^{2 \times 2 }_{\mu \nu }(x) $ possesses the same 
(lattice) symmetry structure than the plaquette, the same symmetry breaking 
terms show up upon an expansion with respect to small lattice spacings. 
One therefore can hope that for a suitable choice of $f$ almost all of the 
breaking terms can be removed. 
Improved actions containing a $ W^{2 \times 2 } $ term were 
considered in~\cite{Beinlich:1995ik,Karsch:1995an} where the prefactor $f$ 
was determined at tree level, i.e., $f_{tree}=1/64$. 
It turned out that already the tree level 
improved action substantially progressed the analysis of the 
hot gluon plasma. Rather than using the tree level improvement, we here 
pursue a non-perturbative improvement: $f$ will be determined by 
minimizing the breaking of rotational symmetry by the underlying lattice. 
Since $O_1$ in (\ref{eq:o1}) explicitly contains terms 
which break rotational symmetry~\cite{Alford:1995hw}, the absence 
of rotational symmetry breaking is directly a measure for 
the suppression of irrelevant terms and, therefore, for improvement. 
\begin{table}[thb]
\begin{ruledtabular}
\caption{ \label{tab:5} Locating $T_c$ for the novel action  
} 
\begin{tabular}{llll}
  $L^3\times N_t$  & $\beta _c$  & $\sigma a^2 (\beta _c) $ & 
  $T_c / \sigma $   \\ \hline 
  $8^3 \times 2$ & $5.718(4)$ & $0.5381(9)$ & $0.682(11) $ \\ 
  $9^3 \times 3$ & $6.21(1) $ & $0.245(2) $ & $0.673(5) $ \\
  $12^3 \times 4$ & $6.55(2)$ & $0.1446(3) $ & $0.657(9) $ \\
  $15^3 \times 5$ & $6.88(2)$ & $0.0961(3)$  & $0.64(1) $ \\ \hline 
\end{tabular}
\end{ruledtabular}
\end{table}
\begin{figure*}[tbh]
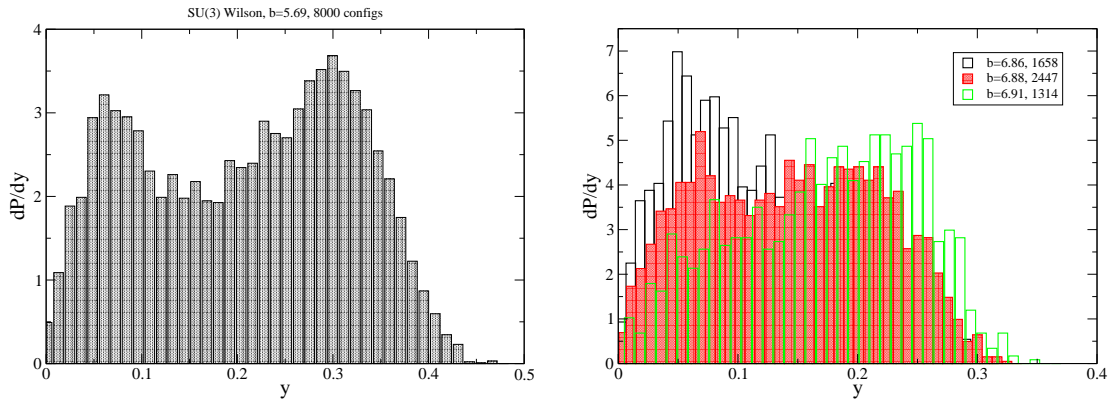

\includegraphics[width=7cm]{wilson_his.eps} 
\hspace{0.4cm}
\includegraphics[width=7cm]{pol.15.5.eps} 
\caption{\label{fig:3} The probability distribution $P(y)$ 
of the Polyakov line close to $T_c$ for the Wilson action (left). $P(y)$ 
for the novel action employing a $15 \times 5$ lattice. 
}
\end{figure*}
\vskip 0.2cm
For an analysis of rotational symmetry breaking, let us 
consider the potential of a static quark-antiquark pair separated 
by a distance vector $\vec{r}$. Choosing one of the main axis of the 
cubic lattice, i.e., $\vec{r} = r \, \hat{e}_i$, $i=1 \ldots 3$, 
yields the ``on-axis'' potential $V_{on}(r)$. Alternatively, one 
might choose $\vec{r} = r /\sqrt{2} \, (1,1,0)^T $ (and permutations) 
and $\vec{r} = r /\sqrt{3} \, (1,1,1)^T $. Contributions from these 
directions to the potential results in the ``off-axis'' potential. 
The numerical method used here to calculate potentials involves 
an overlap enhancement of the Wilson loops with the quark 
ground state along the lines described in~\cite{Bali:ab}. 
In fact, a slightly improved version of overlap enhancement was used, 
the details of which can be found in~\cite{Langfeld:2003ev}. 
Let $\delta V(r)$ denote the statistical error of the potential 
at distance $r$. A measure of rotational symmetry violation 
was constructed in~\cite{deForcrand:1999bi}, i.e., 
\be 
\delta _v^2 \; = \; \sum _{\mathrm{off}} \frac{ 
[ V(r) \; - \; V_{on}(r) ]^2 }{ V(r)^2 \; \delta V ^2 (r) } 
\; \big/ \; \bigl( \sum _{\mathrm{off}} 
\frac{1}{  \delta V ^2 (r) } \bigr) \; . 
\label{eq:del}
\en
In practice, the ``on-axis'' potential is well represented by the fit 
\be 
V_{on}(r) \; = \; a + b/r + c r \; . 
\label{eq:von}
\en 
The latter expression provides 
information on the ``on-axis'' potential at positions $r$ 
which are not multiples of the lattice spacing. 
Figure \ref{fig:1} shows the measure of rotation symmetry breaking, 
$\delta _v$, for several actions. 
Data for $\delta _v$ for the DBW2 and the Wilson case are available 
in~\cite{deForcrand:1999bi}. My results are in good agreement with 
these findings. For the first time, $\delta _v$ for the Iwasaki case 
are reported. We find that Iwasaki's choice for $f$ does not 
completely remove the rotational symmetry breaking effects 
at lattice spacings as large as $a^2 \approx 0.1 \, \mathrm{fm}^2$. 
On coarse lattices, the DBW2 action is less plagued by corrections due 
to irrelevant terms. By contrast, for $a^2 < 0.025\, \mathrm{fm}^2$ 
the Iwasaki action seems to be less biased by lattice artefacts. 
My findings therefore agree with the results of Necco 
in~\cite{Necco:2003vh}. Concerning the $2 \times 2$ action equipped 
with tree level factor $f=1/64$, the improvement on rotational 
symmetry breaking effects is hardly better than in the Wilson case. 

\vskip 0.2cm 
The key idea of non-perturbative improvement is to free the parameter 
$f=1/64$ of the $2 \times 2$ action, 
and to find a choice for $f$ which is optimal with respect 
to the symmetry breaking effects. One possibility for this program 
is to perform a renormalization group study employing lattice perturbation 
theory,  see~\cite{Luscher:1985zq}, or using blocked configurations, 
see~\cite{deForcrand:1999bi}. Here, we adopt a practical point 
of view: For a given $\beta $, we calculate the quantity 
$\delta _v$ for several values $f$. Starting with the tree level 
value $f=1/64$, we systematically increase $f$ in steps of $10^{-3}$, 
and we are monitoring  $\delta _v$. Once the order $a^2$-terms, 
the terms generating the linear slope of $\delta _v$ in figure \ref{fig:1}, 
are eliminated, we stop the procedure. For this aim, $100$ configurations 
were generated using a $10^4$ lattice. Configurations were obtained 
by virtue of the Cabibbo Marinari algorithm~\cite{Cabibbo:zn} supplemented 
with micro-canonical reflections in order to reduce 
autocorrelations (details can be found in~\cite{Langfeld:2003ev}). 
The above procedure for determining $f$ 
is repeated for several values $\beta $ implying that $f$ could 
in principle be a function of $\beta $. However, it turned out that 
the choice 
\be 
f \; = \; 0.056(1) 
\label{eq:f}  
\en 
yields satisfactory results for all $\beta $ values under investigations. 
Note that this value is roughly a factor $3.5$ larger than the tree level 
choice, i.e., $f_{tree}=1/64= 0.015625$. We find at the end that 
the ``$2 \times 2$'' action with $f = 0.056 $ discussed in the present paper 
yields results of the same quality than the DBW2 action, in particular, 
in the regime of coarse lattices. In addition, the novel action seems to be 
comparable with the Iwasaki action at small $a$.

\begin{figure}[tbh]
\vspace{0.5cm}
\includegraphics[width=7cm]{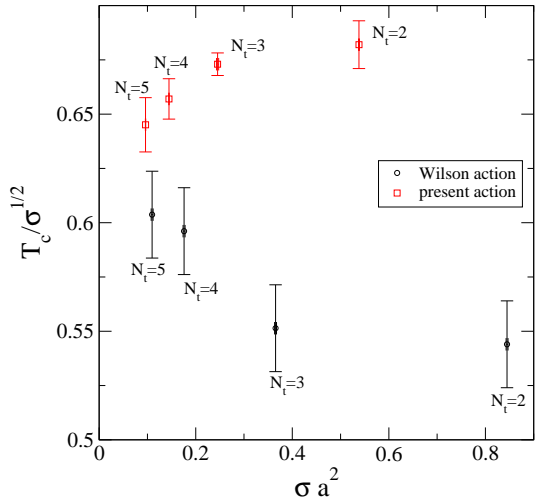} 
\caption{\label{fig:4} Scaling analysis for $T_c/\sqrt{\sigma }$ for 
the Wilson action and the novel action. 
}
\end{figure}
\vskip 0.2cm 
Let us finally explore the scaling properties of the novel action 
for a $16^4$ lattice. 
For each $\beta \in \{ 6.0, \, 6.2, \, 6.4 \}$, 
$136$ independent configurations have been generated. The 
static quark-antiquark potential is fitted with the ansatz 
(\ref{eq:von}). The results for the string tension are listed 
in table \ref{tab:4}. 
Using these values for $\sigma a^2$, we finally express the static 
potential (as well as $r$) 
in physical units ($\sigma = (440 \, \mathrm{fm})^2$ is set 
for the reference scale). The result is shown in figure \ref{fig:2}. 
We observe almost perfect scaling: the data points obtained 
from different $\beta $ values fall on top of a single curve. 
In addition, effects from the breaking of the rotational symmetry 
by the underlying lattice are invisible. 

\vskip 0.2cm 
As a further application, let us study the deconfinement phase transition 
at finite temperatures. The order parameter of the transition 
is the expectation value of the Polyakov line, i.e., 
\be 
y \; := \; \frac{1}{L^3} \left\langle 
\left\vert \sum _{\vec{x}} \frac{1}{3} \tr {\cal P} (\vec{x}) 
\right \vert \right\rangle \; , 
\label{eq:p} 
\en 
where $L$ is the spatial extent of the lattice, and 
$$ 
{\cal P} (\vec{x}) \; = \; \prod _{n=1}^{N_t} U_0(\vec{x},n) 
$$
with $N_t$ being the number of lattice points in Euclidean time 
direction. The temperature $T$ is given by 
$T \; = \; 1/N_t \, a(\beta ) $. 
Since we are only interested in the scaling properties of the 
novel action, we have chosen a physical situation which is accessible 
by a reasonable amount of computer time: Rather than to aspire the 
infinite volume limit $L^3 \rightarrow \infty $, we studied the 
finite volume case $L=3/T_c$. Thereby, $T_c$ is defined as follows: 
Let $P(y)$ denote the probability distribution for the variable 
$y$ in (\ref{eq:p}). At the deconfinement phase transition, 
the probability distribution shows a characteristic double peak 
structure reflecting the first order nature of the transition 
(see figure \ref{fig:3}). 
In the finite volume system, $T_c$ is defined as the temperature at which 
the two peaks of $P(y)$ are of equal height. 
For a given $N_t$, simulations extracted the critical coupling 
$\beta _c$ for the novel action and for the Wilson action 
for comparison. The findings for the novel action are summarized 
in table \ref{tab:5}. For the largest lattice and $\beta =6.91$, 
the auto correlation length was estimated to $\tau \approx 12 $, 
whereas $1300$ configurations were used to estimate $P(y)$. 
Our final results for $T_c/\sqrt{\sigma }$ are summarized 
in figure~\ref{fig:4}. 

In conclusions, a thorough study of the non-perturbatively improved 
$2 \times 2 $ action was performed. While the Iwasaki action shows 
a significant breaking of rotational symmetry for lattice 
spacings larger than $\sigma a^2 > 0.25$, the present action 
yields results in this regime which are even comparable with those of the 
DBW2 action. On the other hand, scaling violations, detected from 
$T_c/\sqrt{\sigma }$, are of the same order as those from 
the Iwasaki action.

\par\bigskip
    
\noindent {\bf Acknowledgments:}
I thank Philippe de Forcrand for useful information.

\end{document}